# RESEARCH OF INTERACTION PROCESSES OF FAST AND THERMAL NEUTRONS WITH SOLUTION OF ORGANIC DYE METHYL ORANGE

Sergey P. Gokov[a], Yuri G. Kazarinov[a,b], Sergiy A. Kalenik[a], Valentin Y. Kasilov[a], Tetiana V. Malykhina[a,b], Yegor V. Rudychev[a,b], Vitaliy V. Tsiats'ko[a,*]

[a]*National Science Center "Kharkiv Institute of Physics and Technology"*
*1, Akademichna str., 61108, Kharkiv, Ukraine*
[b]*Kharkiv V.N. Karazin National University*
*4, Svobody sq., 61022, Kharkiv, Ukraine*
*\*Corresponding Author: stek@kipt.kharkov.ua*

The emergence of powerful sources of ionizing radiation, the needs of nuclear energy, technology and medicine, as well as the need to develop reliable methods of protection against the harmful effects of penetrating radiation stimulated the development of such branches of science as radiation chemistry, radiation biology, radiation medicine. When an organic dye solution is exposed to ionizing radiation, it irreversibly changes color. As a result, the absorbed dose can be determined. The processes of interaction of neutron fluxes with an aqueous solution of an organic dye methyl orange (MO) – $C_{14}H_{14}N_3O_3SNa$, containing and not containing 4% boric acid, have been investigated. The work was carried out on a LINAC LUE-300 at NSC KIPT. A set of tungsten plates was used as a neutron-generating target. The electron energy was 15 MeV, the average current was 20 μA. The samples were located behind the lead shield and without it, with and without a moderator. Using the GEANT4 toolkit code for this experiment, neutron fluxes and their energy spectra were calculated at the location of experimental samples without a moderator and with a moderator of different thickness (1-5 cm). An analysis of the experimental results showed that when objects without lead shielding and without a moderator are irradiated, the dye molecules are completely destroyed. In the presence of lead protection, 10% destruction of the dye molecules was observed. When a five-centimeter polyethylene moderator was installed behind the lead shield, the destruction of dye molecules without boric acid on thermal neutrons was practically not observed. When the fluxes of thermal and epithermal neutrons interacted with a dye solution containing 4% boric acid, 30% destruction of dye molecules was observed due to the exothermic reaction $^{10}B(n, \alpha)$. The research has shown that solutions of organic dyes are a good material for creating detectors for recording fluxes of thermal and epithermal neutrons. Such detectors can be used for radioecological monitoring of the environment, in nuclear power engineering and nuclear medicine, and in the field of neutron capture therapy research in particular.

  The emergence of powerful sources of ionizing radiation, the needs of nuclear energy and technology, as well as the need to develop reliable methods of protection against the harmful effects of penetrating radiation stimulated the rapid development of radiation chemistry, radiation biology, and radiation medicine. At present, intensive development of chemical dosimetry methods is underway. The most convenient model objects for researching the processes of interaction of ionizing radiation with a substance are liquid and solid solutions of organic dyes and pigments [1-14]. Organic dye solutions have intense absorption and fluorescence bands in the visible region of the spectrum. When the dye solution is exposed to ionizing radiation, an irreversible loss of dye color occurs (a decrease in the intensity of the long-wavelength band of the absorption spectrum). In this case, the shape of the absorption spectrum band, as a rule, does not change.

  Previous researches [12-13] have shown that irreversible radiation destruction of dyes in solutions occurs as a result of the oxidation of organic dyes by radicals and radical ions formed during the radiolysis of solvents (OH˙, HO₂˙, etc.). A relatively stable product of the radiolysis of solvents, hydrogen peroxide, also participates in the decolorization of dye solutions.

  The rate of radiation destruction of a dye in a solution essentially depends on both the chemical nature of the dye and the nature and physicochemical properties of the solvent. The lowest radiation resistance of dyes is observed in aqueous solutions, the highest - in solvents whose molecules do not contain oxygen atoms (for example, in dimethylamine), and in solid solutions (for example, in polymer films) [1-14]. So, from the decrease in the intensity of the long-wave absorption band of the dye solution under the action of ionizing radiation, it is possible to determine the value of the radiation dose. Thus, an organic dye solution can serve as a radiation dose detector and be used for radiation monitoring of the environment [2-6]. For example, aqueous solutions of dyes can be successfully used to visually determine the radiation dose in the range of 0.003 – 0.5 Mrad, and polymer films colored with dyes- 0.3 – 40 Mrad.

  According to literature data, the USA and England produce polymethyl methacrylate and paper coated with polyvinyl chloride containing a dye for use in dosimetry. Depending on the degree of degradation of the dye, it is possible to determine doses in the range from 0.1 to 6 Mrad. The US industry produces cellophane films containing some colorants. These films are widely used to measure doses of various types of radiation because they change color when exposed to them. The degree of discoloration is linearly dependent on the dose. In this case, the dose range is from 0.1 to 10 Mrad. All these systems are characterized by the independence of the readings from changes in the dose rate and temperature during irradiation. These systems retain their properties for a sufficiently long period of time after irradiation, which greatly simplifies the measurement process. Before irradiation, they can be stored in the dark for a long time. These systems can be used to determine the spatial distribution of absorbed doses that are received under the influence of high-energy electrons. With their help, the processes of radiation processing of various materials are controlled in production conditions. To solve such problems, a chemical dosimetry method [14] was developed at the Pisarzhevsky Institute of Physical Chemistry of the Academy of Sciences of the Ukrainian SSR, based on the use of films made of colored polyvinyl alcohol. Thus, irradiation of organic dyes can lead to various photochemical reactions. At present, the nature of these processes is still poorly understood.

## FORMULATION OF THE PROBLEM

In this work, the processes of interaction of neutron fluxes with an aqueous solution of an organic dye methyl orange (MO) – $C_{14}H_{14}N_3O_3SNa$, containing and not containing 4% boric acid, have been researched. The work was carried out on a LINAC LUE – 300 at NSC KIPT. The work was carried out with the aim of researching the processes of interaction of neutrons with matter, as well as the possibility of creating detectors based on these materials for registering fluxes of thermal and fast neutrons.

A number of nuclear reactions are used to register neutrons, such as: $^{10}$B (n, α) $^7$Li, $^6$Li (n, α) T, $^3$He (n, p) T, elastic, inelastic scattering, etc. The methods used in the work are elastic neutron scattering and nuclear reaction $^{10}$B (n, α) $^7$Li.

### Elastic neutron scattering

In the case of researching fast neutron fluxes in the elastic scattering reaction, the neutron transfers part of its kinetic energy to the nucleus. The neutron gives up the maximum share of energy in a central (head-on) collision with nuclei. The mass of the neutron differs little from the mass of the proton. A neutron transfers all of its kinetic energy to the nucleus of a hydrogen atom in the event of a central collision. Elastic neutron scattering by hydrogen nuclei is used to register fast neutrons from recoil protons. In this work, an aqueous solution of methyl orange was used. Irreversible radiation destruction of the dye occurred due to the radiolysis of water as a result of oxidation by radicals (OH˙, HO$_2$˙, etc.) and the kinematic destruction of dye molecules in collisions with fast neutrons.

### Nuclear reaction $^{10}$B (n, α) $^7$Li

Natural boron consists of two isotopes: $^{10}$B (18.2%) and $^{11}$B (81.8%). Slow neutrons interact intensely with the nuclei of the $^{10}$B isotope. In the exothermic reaction $^{10}$B (n, α), α-particles and a $^7$Li nucleus appear. This reaction takes place through two channels:

$$^{10}B+n\rightarrow(^7Li)^*+\alpha+E_1;$$

$$^{10}B+n\rightarrow {}^7Li+\alpha+E_2,$$

where $E_1$ and $E_2$ – the reaction energy, is released in the form of the kinetic energy of the reaction products. In the first channel, the $^7$Li nucleus is formed in an excited state with an excitation energy of 0.48 MeV. The excited state of $^7$Li is indicated by an asterisk. The transition of the nucleus from the excited state to the ground state is accompanied by the emission of a γ-quantum with an energy of $E_\gamma$ = 0.48 MeV. Therefore, the first reaction channel can be rewritten as follows:

$$^{10}B + n \rightarrow {}^7Li + \alpha + E_\gamma + E_1.$$

The energy of the $^{10}$B (n, α) $^7$Li reaction is 2.78 MeV. One part of the energy ($E_\gamma$ = 0.48 MeV) is carried by the γ-quantum, the other part ($E_1$ = 2.30 MeV) is released in the form of the kinetic energy of the α-particle and the lithium nucleus. At the same time, the Energy $E_1$ = 2.30 MeV is made up of $E_\alpha$ and $E_{Li}$. The fraction of the α-particle is $E_\alpha$ = 1.47 MeV, and the fraction of the lithium nucleus is $E_{Li}$ = 0.83 MeV. The probability of the reaction proceeding through the first channel is 93% for free neutrons with an energy of about 10 keV. Then this probability gradually decreases, reaching a value of 0.3 at a neutron energy of 1.8 MeV and again increases to 0.5 at a neutron energy of 2.5 MeV. In the second channel, the $^{10}$B (n, α) $^7$Li reaction proceeds with a 7% probability for slow neutrons and, accordingly, with a higher probability for fast neutrons. The reaction energy $E_2$ = 2.78 MeV in this case is completely carried by the α particle and the lithium nucleus.

Thus, when boric acid is added to the dye solution and it is irradiated with a flux of thermal neutrons, the above reactions occur. In this case, the interaction of high-energy α-particles and $^7$Li nuclei with dye molecules leads to their destruction, and the presence of a gamma quantum in the first reaction leads to radiolysis of water and ionization of atoms in molecules.

## EXPERIMENTAL RESEARCHES

The work was carried out at the NSC KIPT on the linear electron accelerator LUE – 300. The electron beam was exposed to a neutron-generating target located at a distance of 40 cm from the output foil. The electron energy was 15 MeV, and the average beam current was 20 μA. The neutron-producing target was a set of tungsten plates. In the immediate vicinity of the target, a lead shield was assembled against scattered electrons and the accompanying gamma background 5 cm thick.

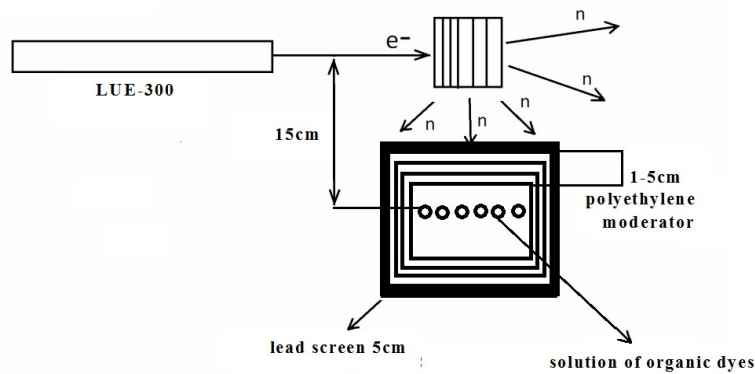

**Figure 1.** Experiment scheme

The schematic of the experiment is shown in Fig. 1. The dye solution in a glass test tube was placed inside a lead shield at a distance of 15 cm from the electron beam axis. The irradiation time was 1 hour, which corresponds to a total neutron flux of $10^{11}$ n/cm$^2$ at the location of the samples. The thickness of the lead shield was 5 cm. Between the lead shield and the test tube, there was a moderator made of polyethylene with a thickness of 1–5 cm.

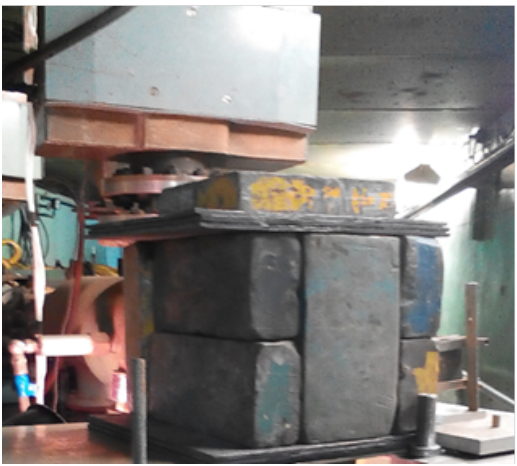

**Figure 2.** Photo of the exit of the LUE – 300 accelerator and the lead shielding of the samples under study.

Figure 2 shows a photograph of the output of the LUE-300 accelerator and the lead shielding of the samples under study. Using the GEANT4 program code [9] for this experiment, the energy spectra of neutron fluxes inside the lead shield at the location of the experimental samples were calculated. The simulation results are shown in Fig. 3. Six cases were considered: 0 - the spectrum of neutrons inside the shield at the location of the sample without a polyethylene moderator, 1–5– spectra of neutrons inside the shield with a gradual increase in the thickness of the moderator from 1 to 5 cm. From Fig. 3 it can be seen that with an increase in the thickness of the moderator layer, the number of fast neutrons decreases significantly (with an increase in the thickness of the moderator from 0 to 5 cm, the number of fast neutrons decreases by a factor of 10). At the same time, the number of thermal neutrons increases insignificantly, approximately 2 times in comparison with the case without a moderator. The total neutron flux for the case without a moderator at the location of the samples for the presented experiment was $10^{11}$ n/cm$^2$.

Several experiments were carried out in the work. In one case, tubes with an aqueous solution of the organic dye methyl orange without boric acid were placed inside and outside the lead shield at a distance of 15 cm from the neutron-producing target.

The main absorption spectra of the irradiated and unirradiated dye are shown in Fig. 4

Fig. 4 that when objects without lead shielding and without a moderator are irradiated, the dye molecules are completely destroyed. This is mostly due to their interaction with scattered electrons and gamma quanta. In the presence of lead shielding, a 10% destruction of dye molecules was observed due to their interaction with fast neutron fluxes.

In another case, one tube with an aqueous solution of an organic dye containing boric acid and the second tube without boric acid were located inside a lead shield in the presence of a polyethylene moderator 5 cm thick. These test tubes were located at a distance of 15 cm from the neutron-forming target. The main absorption spectra of the irradiated and unirradiated dye with and without boric acid are shown in Fig. 5 and 6.

In Fig. 5 and 6, it can be seen that when a polyethylene moderator was installed after a lead shield 5 cm thick and in front of the target, the destruction of dye molecules without 4% boric acid on thermal neutrons was practically not observed. The target consisted of test tubes with solutions of dye (MO) – $C_{14}H_{14}N_3O_3SNa$. When a dye solution containing 4% boric acid interacted with fluxes of thermal and epithermal neutrons, a 30% destruction of dye molecules was observed due to the exothermic reaction 10B (n, α). Researches have shown that solutions of organic dyes, on the one hand, can be a good material for creating detectors of ionizing radiation, in particular, for fluxes of thermal and epithermal neutrons. On the other hand, solutions of organic dyes are a convenient object for studying the processes of interaction of ionizing radiation with matter as a whole.

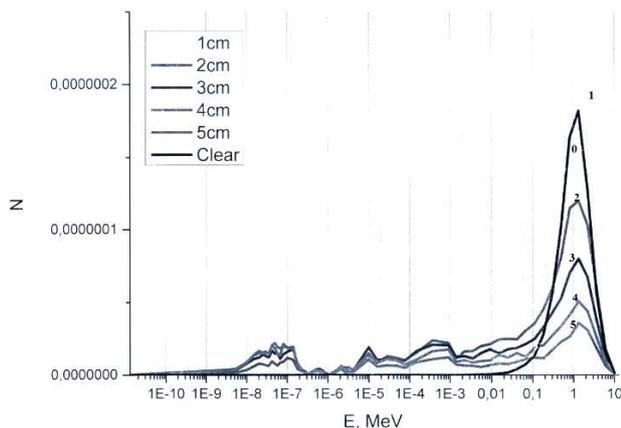

**Figure 3.** Simulation results using the GEANT4 code of the energy spectra of neutron fluxes from a neutron-producing target (electron energy 15 MeV, current 20 μA) with polyethylene moderators of different thicknesses: 1–5 cm and without them.

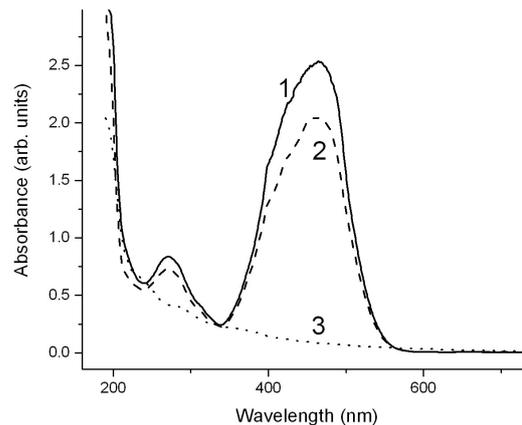

**Figure 4.** Main absorption spectra of irradiated and unirradiated dye. 1 – before irradiation; 2 – irradiated with Pb shield; 3 – irradiated without Pb shield

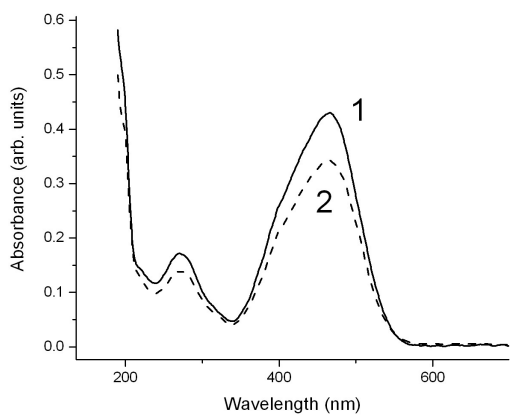

**Figure 5.** Optical absorption spectra of an aqueous solution of a dye before (1) and after (2) irradiation with neutron fluxes in the presence of a polyethylene moderator

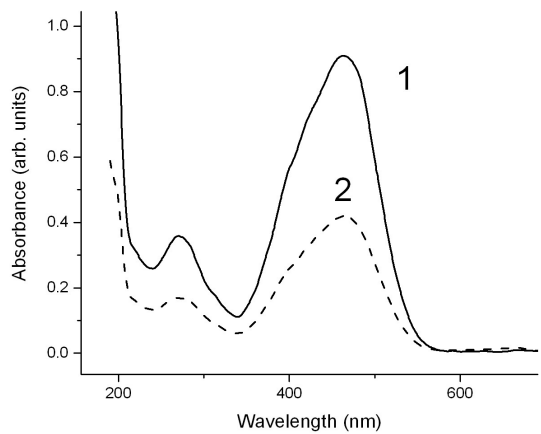

**Figure 6.** Optical absorption spectra of an aqueous solution of a dye with boric acid before (1) and after (2) irradiation with neutron fluxes in the presence of a polyethylene moderator.

## CONCLUSION

In this work, the processes of interaction of neutron fluxes with an aqueous solution of an organic dye methyl orange (MO) – $C_{14}H_{14}N_3O_3SNa$, containing 4% boric acid, and with an aqueous solution of the same dye without boric acid were investigated. In the experimental and simulated parts of the work, these samples were located behind the lead screen and without it. Also, a polyethylene moderator with a thickness of 0 to 5 cm was used in the work.

The energy spectra of neutron fluxes at the location of the experimental samples were calculated using the GEANT4 program code.

An analysis of the experimental results showed that when objects without lead shielding and without a moderator are irradiated, the dye molecules are completely destroyed. This happens mostly due to organic dye interaction with scattered electrons and gamma quanta. In the presence of lead shielding, 10% destruction of dye molecules was observed due to their interaction with fast neutron flux. When a five-centimeter polyethylene moderator was installed behind the lead shield, the destruction of dye molecules without boric acid on thermal neutrons was practically not observed. When the fluxes of thermal and epithermal neutrons interacted with a dye solution containing 4% boric acid, 30% destruction of dye molecules was observed due to the exothermic reaction $^{10}B(n, \alpha)$.

The researches have shown that a solution of organic dye MO is a good material for creating detectors for recording fluxes of thermal and epithermal neutrons, which currently have no analogues. Such detectors can be used for radiation monitoring of the environment, in nuclear power engineering and nuclear medicine, and in the field of neutron capture therapy research in particular.

**ORCID IDs**
**Sergey P. Gokov**, https://orcid.org/0000-0002-3656-3804; **Yuri G. Kazarinov**, https://orcid.org/0000-0001-5143-8545
**Sergiy A. Kalenik**, https://orcid.org/0000-0002-1460-5081; **Valentin I. Kasilov**, https://orcid.org/0000-0002-1355-311X;
**Tetiana V. Malykhina**, https://orcid.org/0000-0003-0035-2367; **Yegor V. Rudychev**, https://orcid.org/0000-0002-1453-2062
**Vitaliy V. Tsiats'ko**, https://orcid.org/0000-0002-7347-0500